\input{aipcheck}

\documentclass[
    final           % use final for the camera ready runs
  ]
  {aipproc}

\layoutstyle{6x9}

\begin{document}

\title{Building and Destroying Symmetry in 1-D Elastic Systems}

\classification{43.35.+d,63.20.Pw,43.40.Cw}
\keywords{Vibrating rods, Anderson localization, Wannier-Stark ladders, Symmetry breaking}

\author{J.~Flores}{
  address={Instituto de F\'{\i}sica, Universidad Nacional Aut\'onoma de M\'exico, A. P. 20-364, 01000, M\'exico, D.~F., M\'exico}
}

\author{G.~Monsivais}{
  address={Instituto de F\'{\i}sica, Universidad Nacional Aut\'onoma de M\'exico, A. P. 20-364, 01000, M\'exico, D.~F., M\'exico}
}

\author{P.~Mora}{
  address={Instituto de F\'{\i}sica, Universidad Nacional Aut\'onoma de M\'exico, A. P. 20-364, 01000, M\'exico, D.~F., M\'exico}
}

\author{A.~Morales}{
  address={Instituto de Ciencias F\'{\i}sicas, Universidad Nacional Aut\'onoma de M\'exico, A.P. 48-3, 62251, Cuernavaca, Morelos, M\'exico}
}

\author{R.~A.~M\'endez-S\'anchez}{
  address={Instituto de Ciencias F\'{\i}sicas, Universidad Nacional Aut\'onoma de M\'exico, A.P. 48-3, 62251, Cuernavaca, Morelos, M\'exico}
}

\author{A.~D\'{\i}az-de-Anda}{
  address={Instituto de Ciencias F\'{\i}sicas, Universidad Nacional Aut\'onoma de M\'exico, A.P. 48-3, 62251, Cuernavaca, Morelos, M\'exico}
}

\author{L.~Guti\'errez}{
  address={Instituto de Ciencias F\'{\i}sicas, Universidad Nacional Aut\'onoma de M\'exico, A.P. 48-3, 62251, Cuernavaca, Morelos, M\'exico}
}

\begin{abstract}
Locally periodic rods, which show approximate invariance with respect to translations, are constructed by joining $N$ unit cells. The spectrum then shows a band spectrum.
%According to whether the unit cell consists of one or two elements, the resulting normal-mode spectrum is a band spectrum or one formed by acoustical and optical spectra. 
We then break the local periodicity by including one or more defects in the system. When the defects follow a certain definite prescription, an analog of the Wannier-Stark ladders is gotten; when the defects are random, an elastic rod showing Anderson localization is obtained. In all cases experimental values match the theoretical predictions.
\end{abstract}

\maketitle

%%%%%%%%%%%%%%%%%%%%%%%%%%%%%%%%%%%%%%%%%%%%
%% MAINMATTER
%%%%%%%%%%%%%%%%%%%%%%%%%%%%%%%%%%%%%%%%%%%%

{\em \noindent We dedicate this paper to Marcos Moshinsky, who was always in love with symmetry}

\section{Introduction}

Due to similarities between the stationary Schr\"odinger equation and the time independent wave equation, several classical analogs of quantum systems have been analyzed in the last few years. Among them, mechanical~\cite{Maynard}, microwave~\cite{Stockmann}, optical~\cite{Agarwal,Ghulinyan,Monsivaisetal1990,DelCastillo}, and elastic systems~\cite{Moralesetal2002} have been studied along these lines. This approach is sometimes, jokingly speaking, referred to as {\em megascopic physics}.

\begin{figure}[h!]
 \includegraphics[width=0.8\columnwidth]{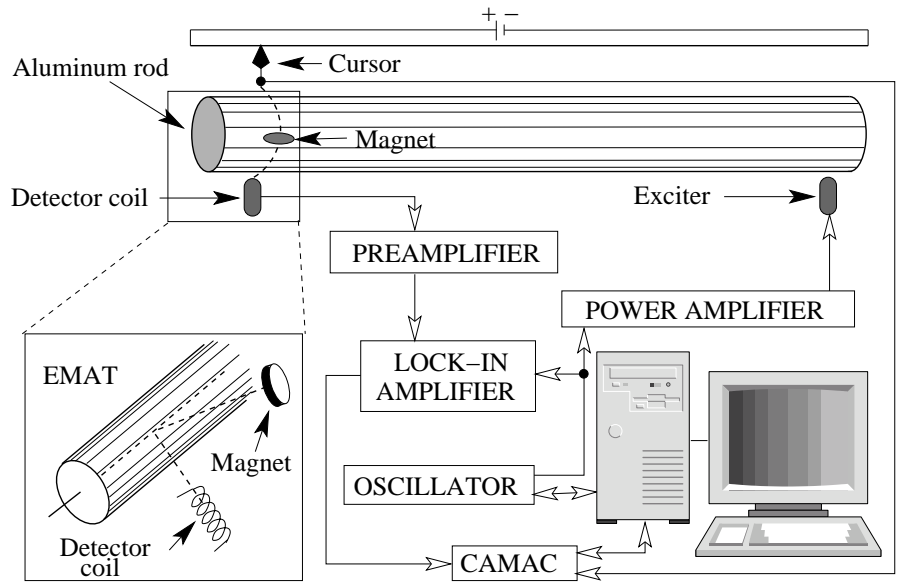}
 \caption{Experimental setup to measure the elastic vibrations of an aluminum rod. In the inset the EMAT is configured for torsional waves. The exciter should be configurated in the same way.}
 \label{Fig.ExperimentalSetup}
%-------------------------------------------FIGURE 1
\end{figure}

\begin{figure}[h!]
 \includegraphics[width=0.8\columnwidth]{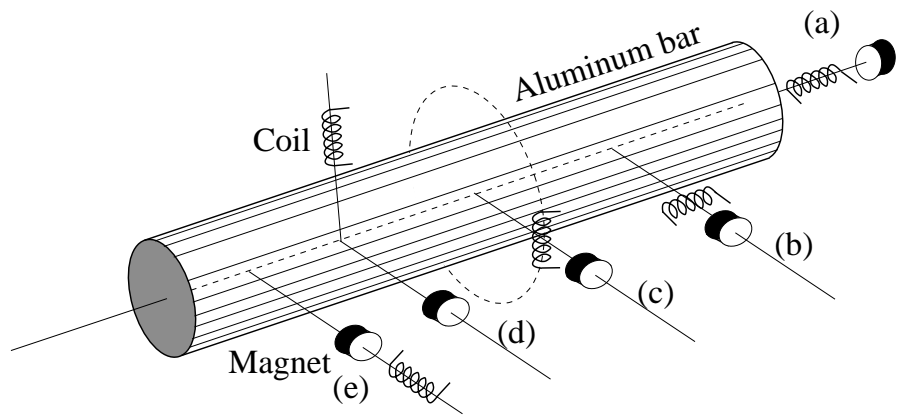}
 \caption{Different configurations of the permament magnet and of the coil, excite different modes: (a) y (b) compressional; (c) y (d) torsional; (e) flexural}
 \label{Fig.EMATConfigurations}
%-------------------------------------------FIGURE 2
\end{figure}

In this paper we shall present a review of some analogs that we have recently discussed using one-dimensional elastic rods. Normal-mode amplitudes and frequencies, which are the classical analogs of quantum wave functions and eigenenergies, respectively, are obtained experimentally and compared with numerical calculations. The experimental set up is shown in Fig.~\ref{Fig.ExperimentalSetup} where an electromagnetic acoustic transducer (EMAT), consisting of a permanent magnet and a coil, is used to excite and detect the vibrations of the rod. The transducer operates through the interaction of eddy currents in the metallic sample with a permanent magnetic field and an oscillating one. This EMAT is very flexible and selective; it can excite or detect either compressional, torsional, or flexural vibrations according to the relative position of the permanent magnet and the coil, as shown in Fig.~\ref{Fig.EMATConfigurations}. Experimental details are discussed in Refs.~\cite{Moralesetal2002,Moralesetal2001}.

\begin{figure}[h!]
 \includegraphics[width=0.8\columnwidth]{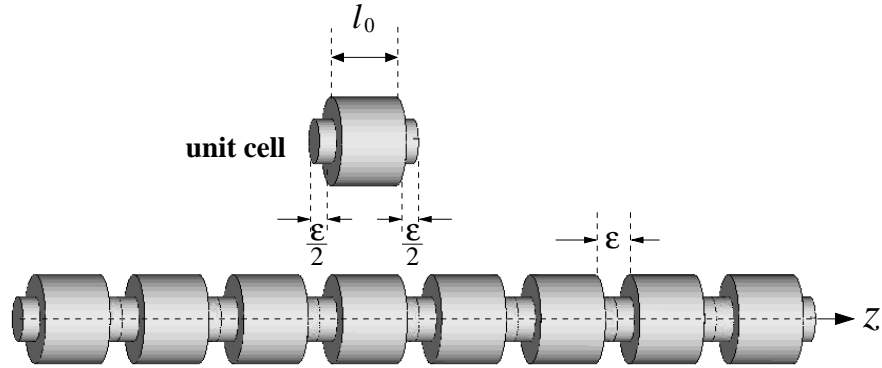}
 \caption{Locally periodic rod. The unit cell consists of a small cylinder of length $l_0$ and two smaller cylinders of length $\epsilon/2$.}
 \label{Fig.LPR}
%-------------------------------------------FIGURE 3
\end{figure}

We shall first consider the building up of a rod which shows local translational invariance. Since we construct it with a finite number $N$ of unit cells (see Fig.~\ref{Fig.LPR}), we refer to it as a locally periodic rod~\cite{GriffithsSteinke}. A band spectrum emerges when $N$ grows. This result is also valid for flexural vibrations, which obey instead a fourth order differential equation~\cite{Graff,DiazDeAndaetal}. If the unit cell consists of two small different rods, an elastic analog of the diatomic chain results: optical and acoustical bands are formed as $N$ grows.

We then induce a symmetry breaking by changing the length $l_0$ of one of the central small rods that form the locally periodic system of Fig.~\ref{Fig.LPR}. A normal-mode frequency is found in the forbidden band and corresponds to a localized state. The number of rods with different lengths is then increased until all of them are distinct. We proceed in two different manners. If the lengths of the rods are altered following a systematic law, the analog of the Wannier-Stark ladders can be obtained. If the lengths are selected as random numbers, the elastic analog of Anderson localization is achieved.

In what follows we shall present our results with the aid of a set of figures, omitting the details, which the interested reader can find in the references given to our published work.

\section{BUILDING TRANSLATION SYMMETRY}

\begin{figure}[h!]
 \includegraphics[width=0.8\columnwidth]{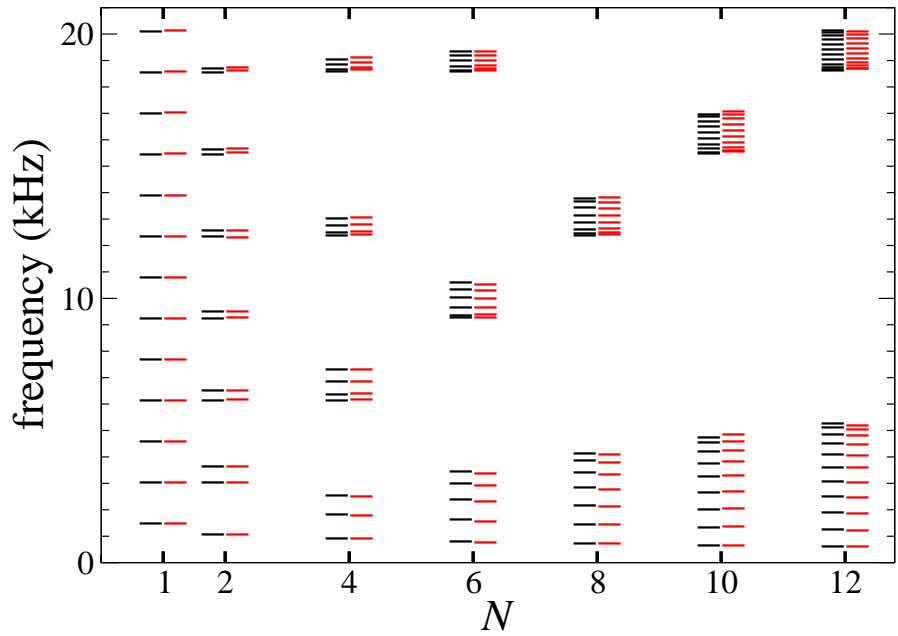}
 \caption{Spectrum of compressional waves of a locally periodic rod. As the number of unit cells increases, a band spectrum emerges.}
 \label{Fig.BandEmergence}
%-------------------------------------------FIGURE 4
\end{figure}

\begin{figure}[h!]
 \includegraphics[width=0.8\columnwidth]{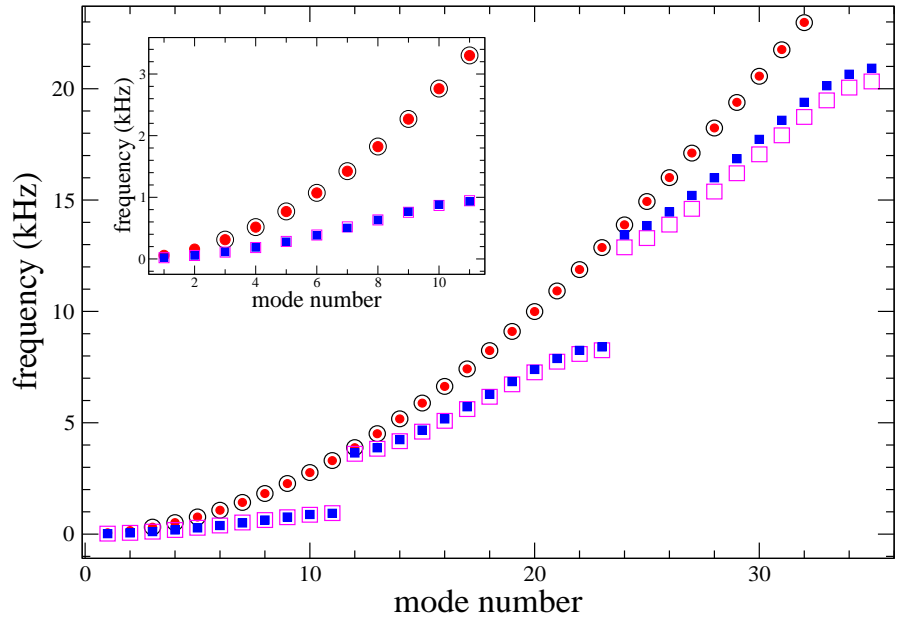}
 \caption{Bloch theorem applies independently of the structure of the Lagrangian. Band spectra are also observed for torsional waves and even for flexural vibrations. Here we show the result for flexural vibrations. The circles correspond to the uniform rod, experiment empty circles, Timoshenko's beam theory is given by the filled circles. The squares show the results for the locally periodic rod, 12 cells, experiment empty squares, Timoshenko's beam theory is given by the filled squares.}
 \label{Fig.FlexionalBandSpectrum}
%-------------------------------------------FIGURE 5
\end{figure}

In Fig.~\ref{Fig.BandEmergence} we present the experimental spectrum of compressional waves of a rod as the number $N$ of unit cells is increased. When $N$ is large enough the spectrum consists of allowed and forbidden bands, each consisting of $N$ levels. The wave amplitudes are also obtained and compared with the ones obtained numerically using the transfer matrix method~\cite{Moralesetal2002}. The waves are extended along the rod. This is consistent with Bloch's theorem and what was found numerically for a Kronig-Penney model~\cite{FloresMonsivais} or for a number of other systems~\cite{GriffithsSteinke}. The same result holds for torsional vibrations~\cite{Moralesetal2002} and, even more, for flexural vibrations~\cite{DiazDeAndaetal}. This is interesting in itself since the latter obey an equation which is not of the Helmholtz type, but rather a fourth order differential equation, either the Bernoulli-Euler one or the more accurate one at higher modes, the so called Timoshenko equation~\cite{Graff}. The band spectrum for flexural vibration is shown in Fig.~\ref{Fig.FlexionalBandSpectrum}.

\section{DESTROYING THE SYMMETRY}

\begin{figure}[h!]
 \includegraphics[width=0.8\columnwidth]{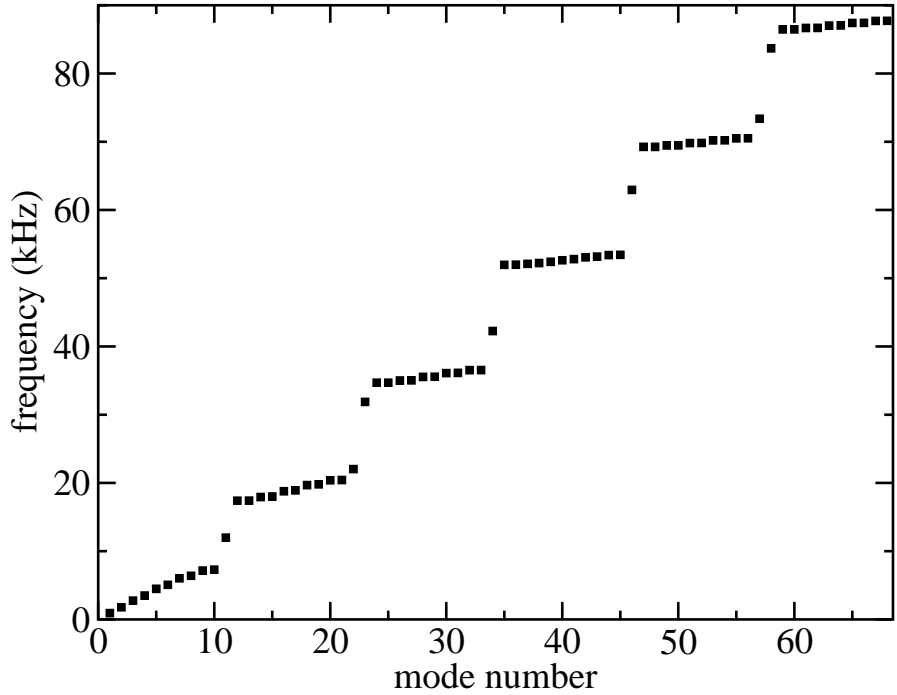}
 \caption{One or more normal-mode frequencies appear in the forbidden gap. %Note the agreement between theory and experiment.
}
 \label{Fig.TopologicalDefectSpectrum}
%-------------------------------------------FIGURE 6
\end{figure}

\begin{figure}[h!]
 \includegraphics[width=0.8\columnwidth]{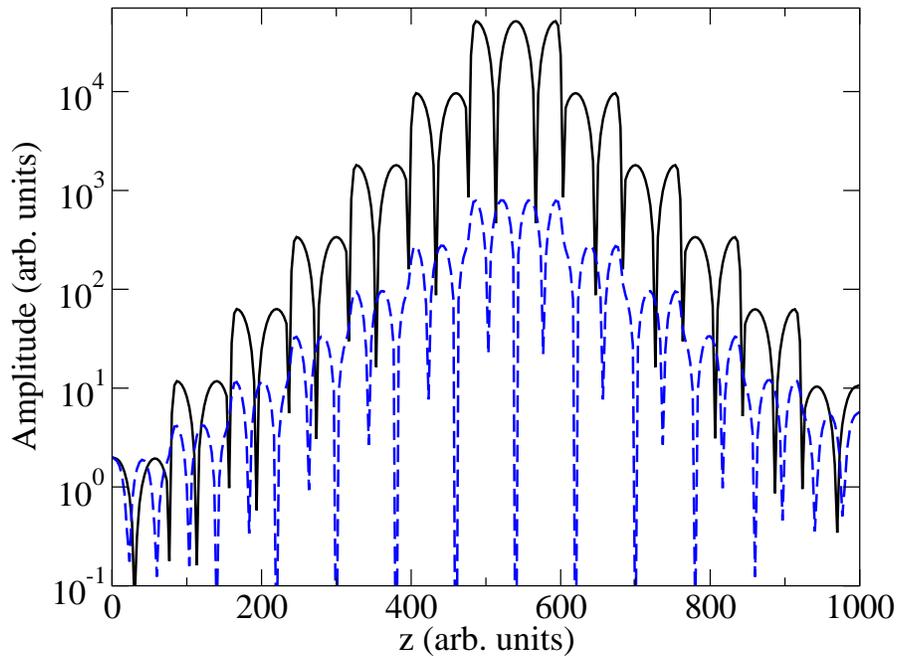}
 \caption{The states in the forbidden band are localized at the topological defect. Mode 24 continuous line, and mode 25, dashed line.}
 \label{Fig.TopologicalWA}
%-------------------------------------------FIGURE 7
\end{figure}

We start by introducing in the locally periodic system depicted in Fig.~\ref{Fig.LPR} a single small rod with a different length $l = l_0 + h$. This is referred to in solid state physics as a topological defect, since it destroys the long range order of the locally periodic or symmetrical rod. We have measured and calculated the spectrum for several values of $h$~\cite{Moralesetal2003}. As given in Fig.~\ref{Fig.TopologicalDefectSpectrum}, there appears a frequency in the first forbidden band of the ordered rod. This is consistent with what one learns from the band theory of crystals. However, in the second forbidden gap two levels lie. In some of the higher gaps two frequencies are also found.
%
%{\bf However, in the same figure it will be seen that at higher frequencies more than one normal mode can have a frequency lying in the second forbidden band. As a matter of fact, for $h$ large enough three or more states can move to the frequency gap.} 
%
These state are localized, as seen is Fig.~\ref{Fig.TopologicalWA}. The existence of such localization was to be expected from the independent rod model because, if there exists a normal-mode frequency proportional to $(l_0 + h)^{-1}$, the neighboring rods of the defect do not resonate.

We can now go on introducing more defects in the locally periodic rod. We first do it in an ordered way to simulate, for example, the effect of an external constant static electric field on a charged quantum particle that is moving in a periodic potential. This is the case, indeed, with the Wannier-Stark ladders (WSL). In 1960, Wannier~\cite{Wannier} suggested that a static external electric field would destroy the electron band spectrum and instead produce a picket fence one, in which the spacing between the energy levels is a constant, proportional to the electric field and the lattice constant. The WSL in solid state were finally observed decades later using superlattices~\cite{Superlattices1}. They have recently been found in optics as well~\cite{Agarwal,Ghulinyan}. As will be shown, by modifying appropriately the small rods in a systematic way one can obtain an elastic analog of WSL \cite{DiazDeAndaetal}.

\begin{figure}[h!]
 \includegraphics[width=0.8\columnwidth]{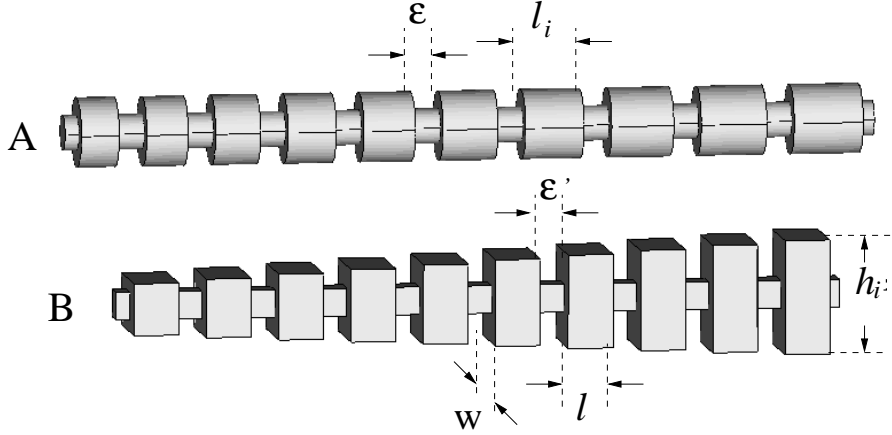}
 \caption{Elastic rods showing the Wannier-stark ladders}
 \label{Fig.WSLRods}
%-------------------------------------------FIGURE 8
\end{figure}

Let us first use an independent rod model. The normal-mode torsional frequencies $f_j^i$ of a rod of length $l_i$ and wave velocity $c_i$ are given by the well known expression~\cite{Graff}
\begin{equation}
 f_j^i=\frac{c_i}{2 l_i} j,
\end{equation}
where $j$ is the number of nodes in the wave amplitude. We consider two options: either $l_i = l_0 / (1+i\gamma)$ and $c_i = \sqrt{G/\rho}$, with $\gamma$ an adimensional parameter, or $l_i = l_0$ and $c_i = c(1+i\gamma)$. This happens for systems A or B, respectively, which are shown in Fig.~\ref{Fig.WSLRods}~\cite{Gutierrezetal}. Here $\rho$ is the mass density and $G$ the shear modulus of the rods and $c$ the wave velocity. 

In the first case, which is the elastic analog of the optical system discussed in Ref.~\cite{Agarwal}, we have
\begin{equation}
 f_j^i=\frac{l_0}{2}\sqrt{\frac{G}{\rho}} \left( 1+ i \gamma \right) j
\end{equation}
and the difference $\Delta f_j^i=f_j^{i+1}-f_j^{i}$ is equal to
\begin{equation}
 \Delta f_j^i=\frac{n}{2L}\sqrt{\frac{G}{\rho}} j \gamma =\Delta f_j =j \Delta_1
\end{equation}
which is independent of the index $i$.

%In the second case B, which is the elastic analog of the optical system dealt with in Ref.~\cite{Ghulinyan}, the velocity $c_i$ is given by
%
%\begin{equation}
% c_i=\frac{G \alpha_i}{\rho I_i}
%\end{equation}
%
%where $I = (hw^3+h^3 w)/12$ is the moment of inertia and $\alpha$ is given by~\cite{Navier}
%
%\begin{equation}
% \alpha \left( h,w \right)=\frac{256}{\pi^2}\sum_{m=0}^\infty \sum_{p=0}^\infty 
%  \frac{1}{\left( 2m+1 \right)^2\left( 2p+1 \right)^2} 
%  \frac{h w}{\left( \frac{2m+1}{h} \right)^2+\left( \frac{2p+1}{w} \right)^2}
%\label{Eq.Navier}
%\end{equation}
%
%For beams of varying $h_i$ we have verified these expressions experimentally~\cite{Monsivaisetal}; see Fig.~\ref{Fig.Navier}. Using Eq.~\ref{Eq.Navier} we have obtained values of $h_i$ such that $c_i = i c$ with $c$ given by $(G \alpha/I_1) 1/2$ and construct therefore system B. 

%\begin{figure}[h!]
% \caption{Experimental verification of the Navier formula (1827)}
% \label{Fig.Navier}
%\end{figure}

\begin{figure}[h!]
 \includegraphics[width=0.8\columnwidth]{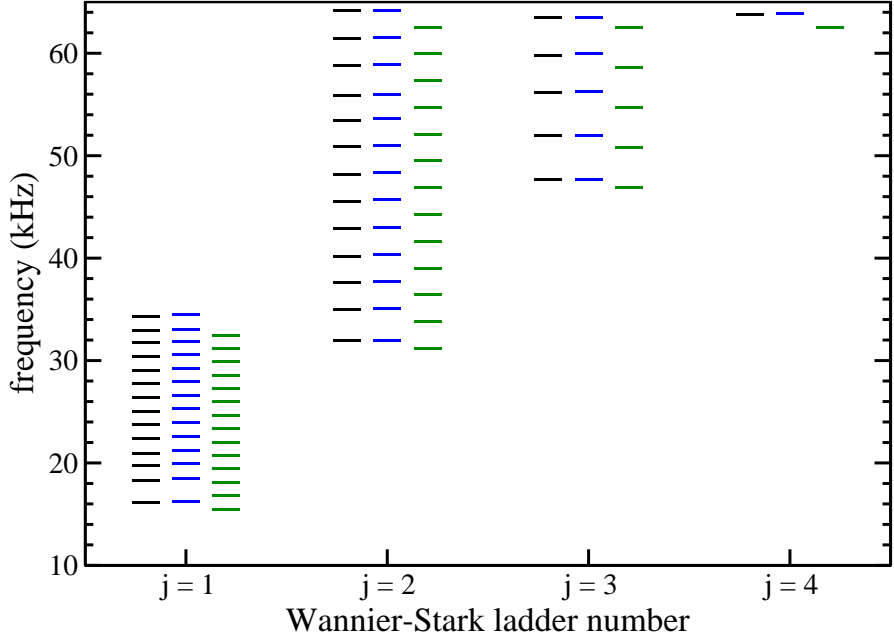}
 \caption{Wannier-Stark ladder spectrum for system A}
 \label{Fig.WSSpectrumSystemA}
%-------------------------------------------FIGURE 9
\end{figure}

\begin{figure}[h!]
 \includegraphics[width=0.8\columnwidth]{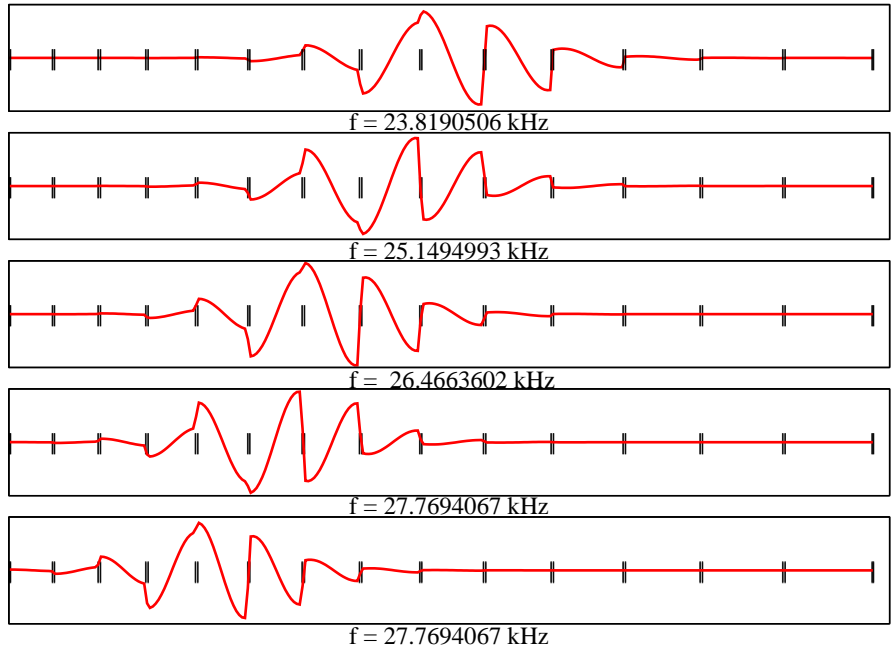}
 \caption{Wannier-Stark ladder wave amplitudes obtained with the transfer matrix method.}
 \label{Fig.WSLWA}
%-------------------------------------------FIGURE 10
\end{figure}

\begin{figure}[h!]
 \includegraphics[width=0.8\columnwidth]{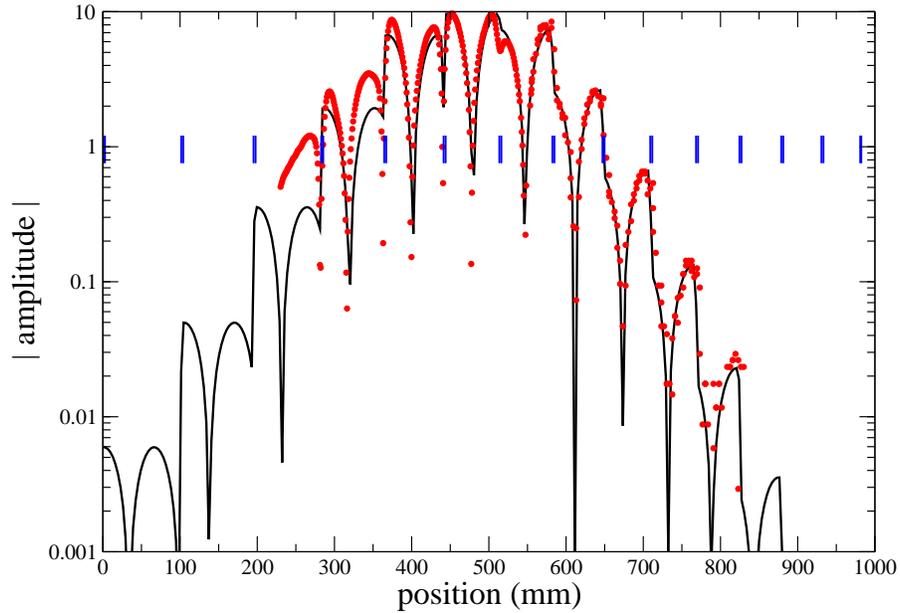}
 \caption{Experimental wave amplitude of the Wannier-Stark ladder and its comparison with the theory.}
 \label{Fig.WSLExperimentalWA}
%-------------------------------------------FIGURE 11
\end{figure}

We should remark that in both cases $\Delta_i^j$, is independent of $i$, a fact which will lead to a WSL as we will now show. In Fig.~\ref{Fig.WSSpectrumSystemA} the WSL measured for system A are compared with the numerical values obtained using the transfer matrix method. The normal-mode amplitudes are localized as one would expect from the independent rod model. This is shown in Fig.~\ref{Fig.WSLWA} for several modes in a given ladder; experimental wave amplitudes fit very well the theoretical ones, as exemplified in Fig.~\ref{Fig.WSLExperimentalWA}. The same conclusions can be obtained by analyzing system B; results for system B can be found in References~\cite{Gutierrezetal,Monsivaisetal}.

A very different effect appears if one now introduces random defects: a disordered rod is built. This is reminiscent of the well known Anderson localization. The quantum mechanical states in a disordered chain of potentials show always a localization length $\xi$. Let us now consider what happens in the elastic case, when we analyze a 1-D system formed by $N$ small rods each of length $l_i$, with $l_i = l_0 +  \epsilon_i$ and $\epsilon_i$ an independent random variable distributed according to $P(\epsilon_i) = 1/2$ for $-1\le \epsilon_i \le 1$ and $P(\epsilon_i)= 0$ otherwise. The disorder is an uncorrelated one in this case. 

\begin{figure}[h!]
 \includegraphics[width=0.8\columnwidth]{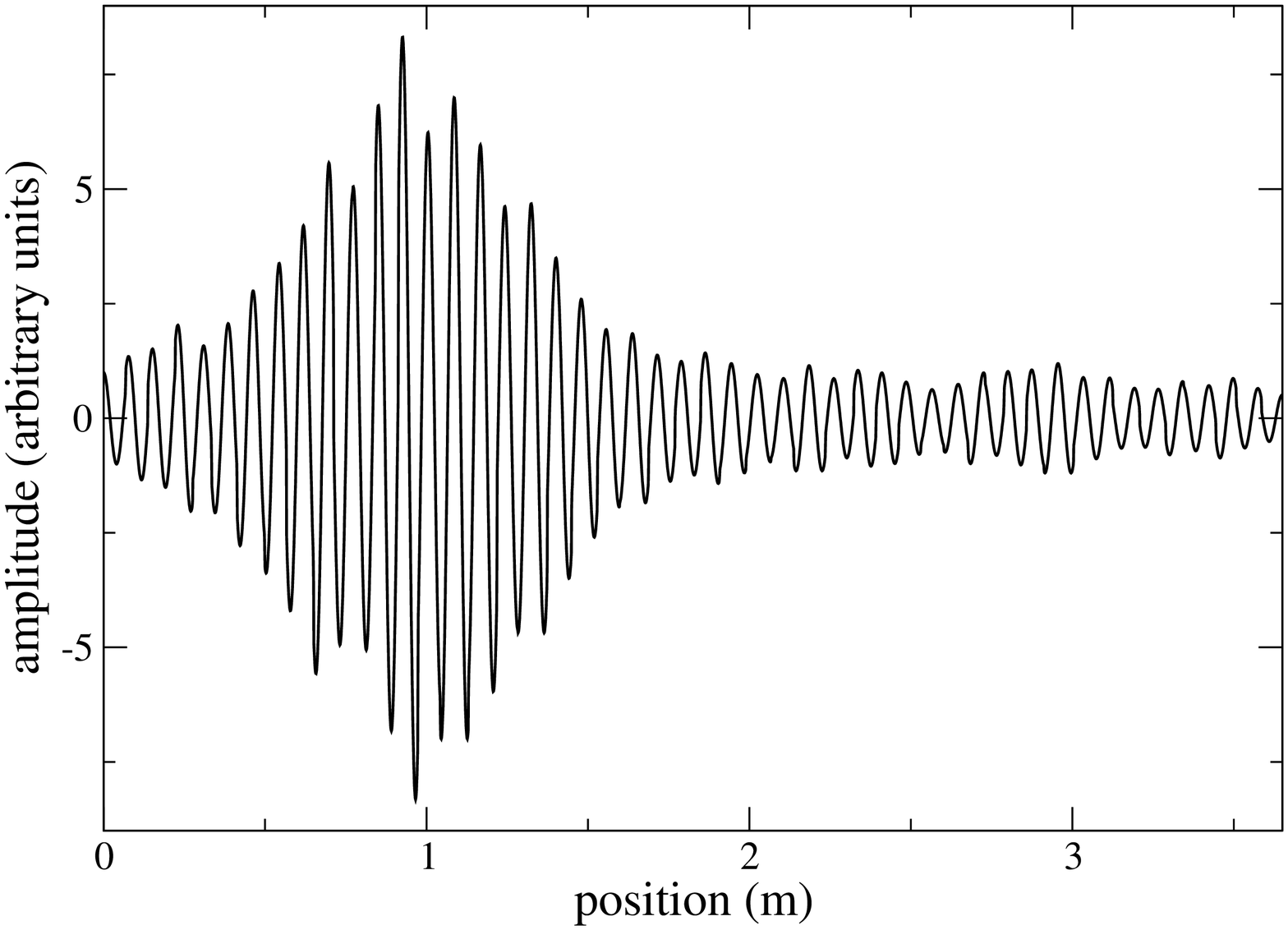}
 \caption{Localized wave function calculated with the method of the Poincar\'e map of a disordered rod.}
 \label{Fig.LocalizedWA}
%-------------------------------------------FIGURE 12
\end{figure}

\begin{figure}[h!]
 \includegraphics[width=0.8\columnwidth]{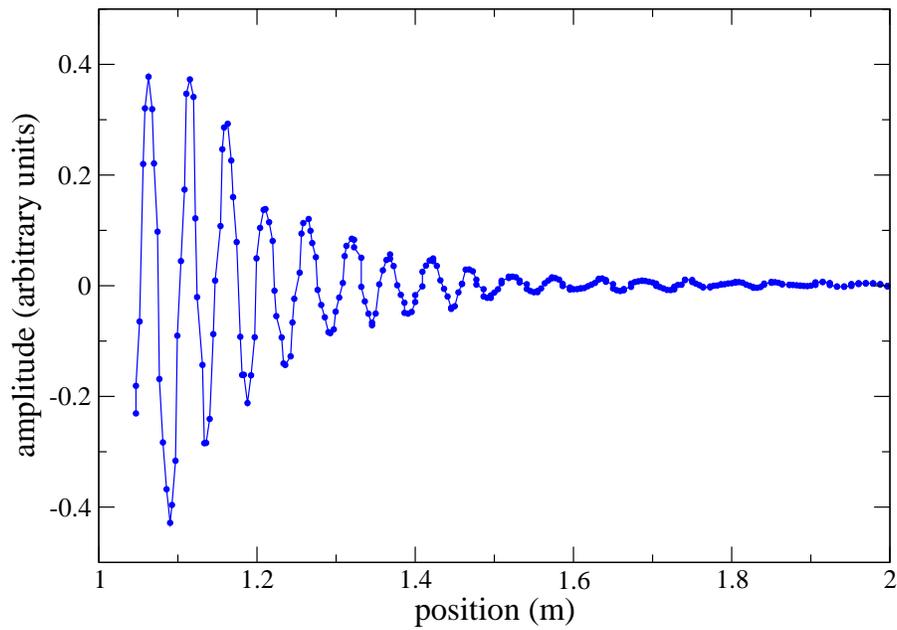}
 \caption{Measured localized wave amplitude of a disordered rod.}
 \label{Fig.ExperimentalLocalizedWA}
%-------------------------------------------FIGURE 13
\end{figure}

\begin{figure}[h!]
 \includegraphics[width=0.8\columnwidth]{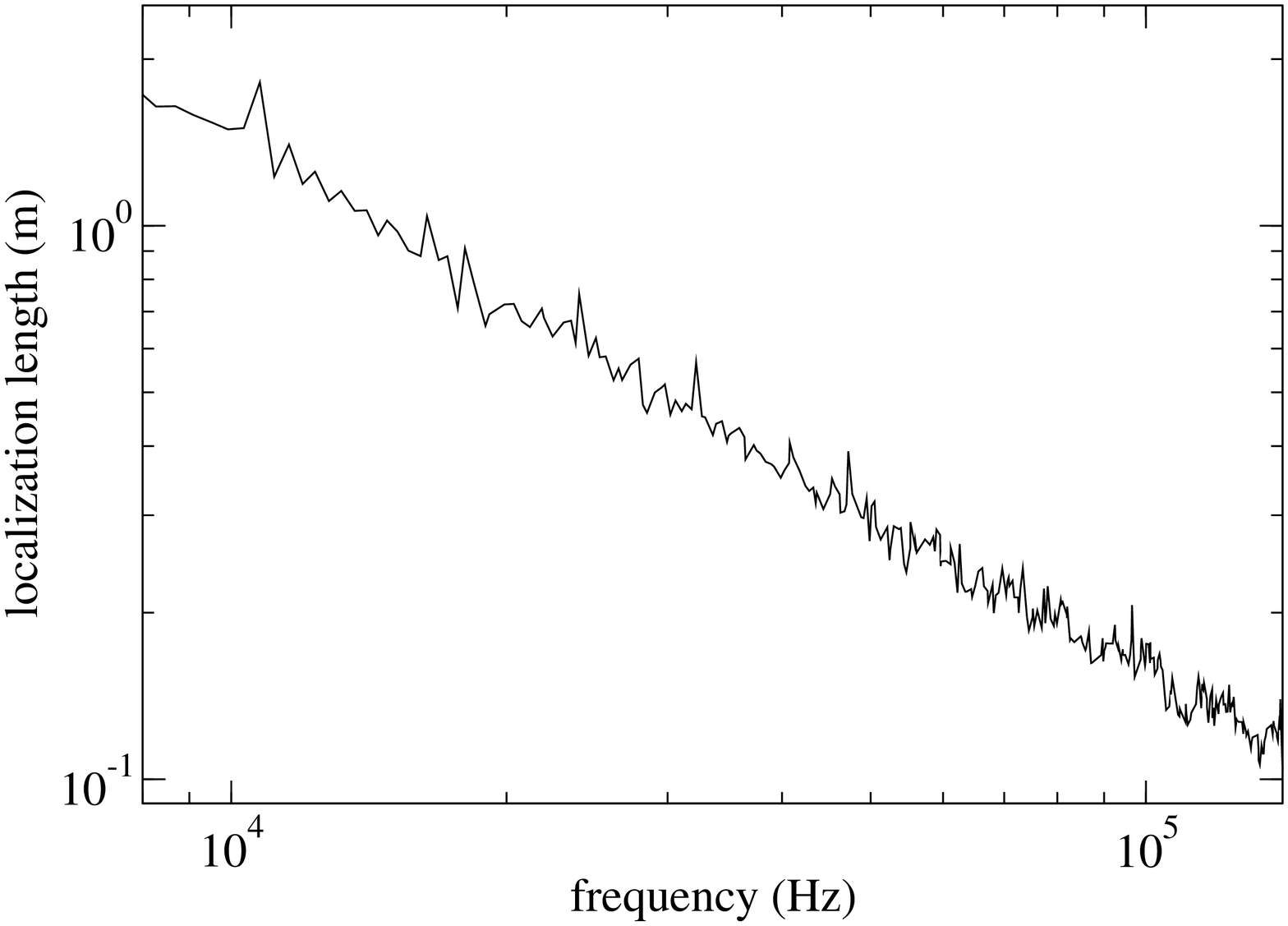}
 \caption{Localization length as a function of frequency}
 \label{Fig.Localization}
%-------------------------------------------FIGURE 14
\end{figure}

We first compute the normal-mode amplitudes. It is found indeed that the Anderson transition takes place, the wave functions being localized, as the examples given in Fig.~\ref{Fig.LocalizedWA} and~\ref{Fig.ExperimentalLocalizedWA} show. Nevertheless, the average localization varies with frequency; see Fig.~\ref{Fig.Localization}. One can also study the statistical properties of the disordered rod spectrum. Using, as is common in quantum chaos~\cite{Brodyetal}, the nearest-neighbor spacing distribution, we find that this distribution also changes with frequency in the elastic case. To characterize it we use the Brody repulsion parameter~\cite{Brody}:
\begin{equation}
 P_\beta(z)=A z^\beta \exp(-\alpha z^{\beta+1});\qquad z=s/d
\end{equation}
with $A=\beta+1$, $\alpha=\left[ \Gamma \left( \beta+ \frac{2}{\beta}+1\right)\right]^{\beta+1}$, $s$ the space between frequencies and $d$ the mean level spacing. When $\beta=0$ a Poisson spectrum is dealt with, for $\beta=1$ the Wigner distribution, which shows level repulsion, is obtained; the Gaussian orthogonal ensemble (GOE) corresponds to the value $\beta=0.953$.

\begin{figure}[h!]
 \includegraphics[width=0.8\columnwidth]{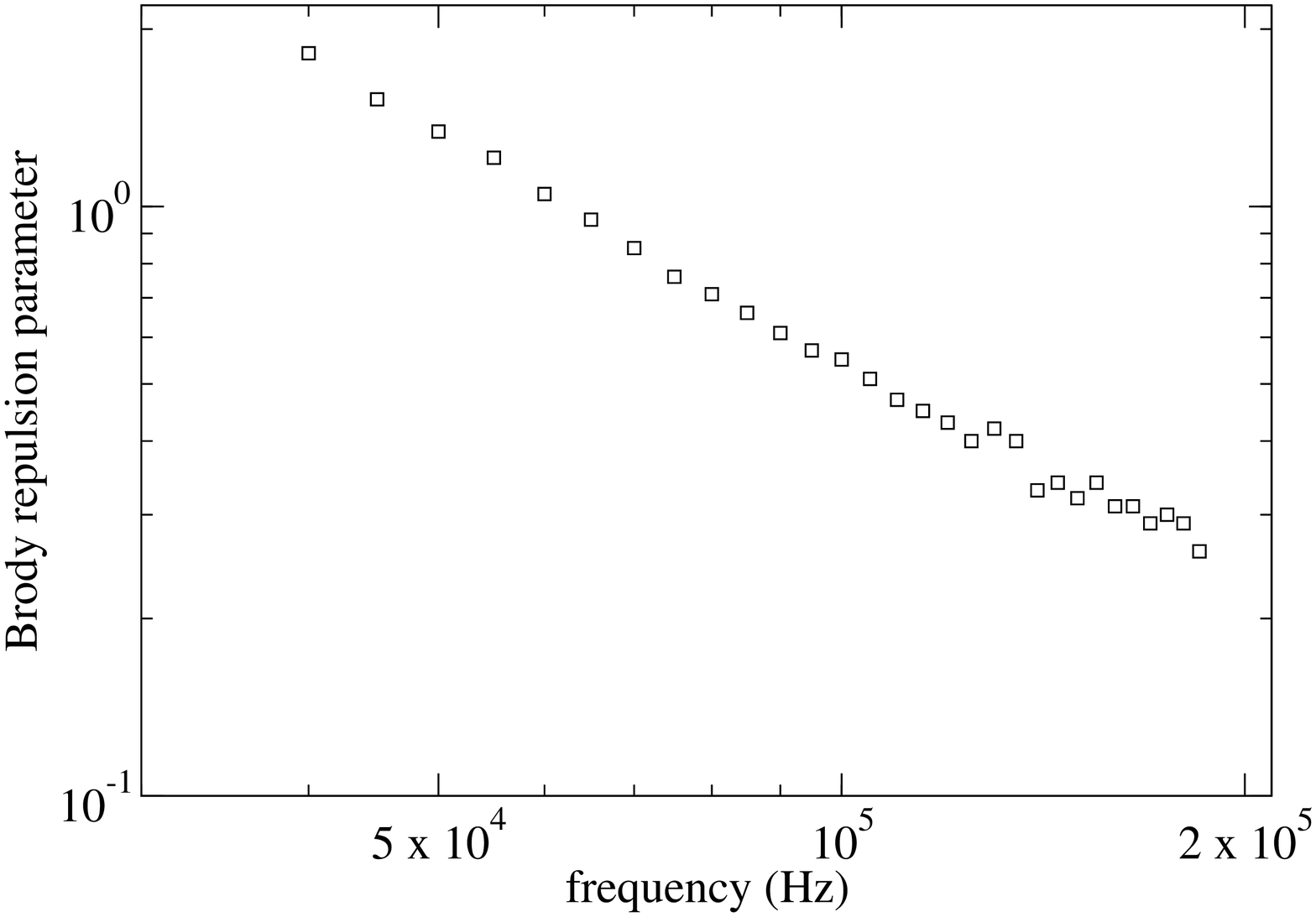}
 \caption{Repulsion parameter as a function of frequency}
 \label{Fig.RepulsionParameter}
%-------------------------------------------FIGURE 15
\end{figure}

\begin{figure}[h!]
 \includegraphics[width=0.8\columnwidth]{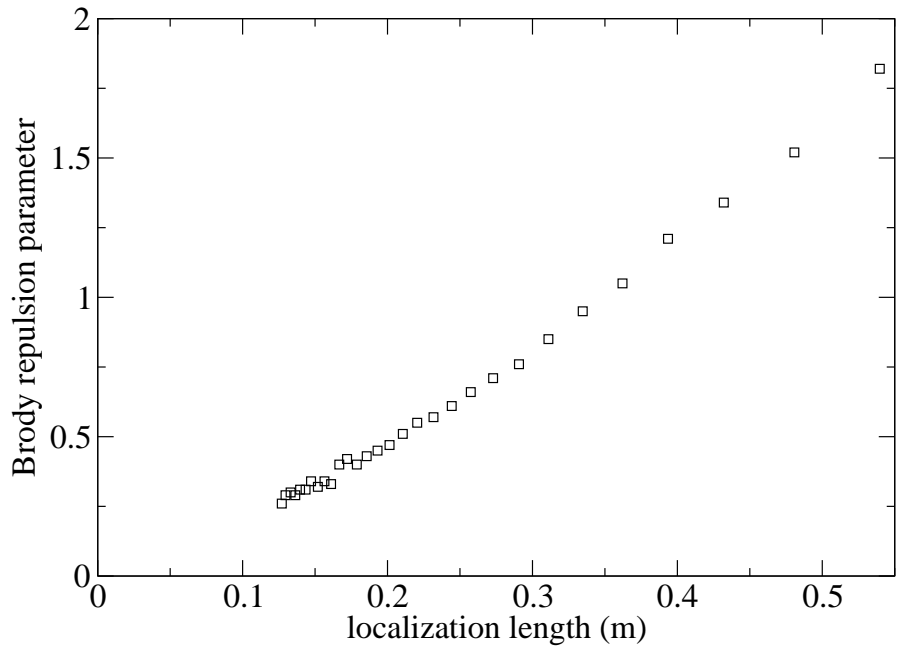}
 \caption{Repulsion parameter vs localization length}
 \label{Fig.BetavsXi}
%-------------------------------------------FIGURE 16
\end{figure}

For the disordered rod the parameter $\beta$ is a function of frequency, showing the richness of elastic systems; this is shown in Fig.~\ref{Fig.RepulsionParameter}. If one eliminates the frequency, Fig.~\ref{Fig.BetavsXi} is obtained, showing that the repulsion parameter is to a good approximation a linear function of the localization length. A similar result was obtained numerically using a tight-binding Hamiltonian with nearest neighbor interaction and diagonal disorder~\cite{Izrailev}.

\section{Conclusions}

We have discussed, both from experimental and numerical points of view, different configurations of elastic rods. They serve as classical analogs of 1-D quantum systems. We first build up a locally periodic rod and a band spectrum emerges. %From this rod another one is built which can be considered the analog of a diatomic chain: the acoustical and optical bands emerge. 
We then destroy the local translational symmetry by introducing defects in the rod. If one does it following specific rules, the analog of Wannier-Stark ladders can be obtained. If the defects have random properties, features of the Anderson transition take place. In all cases, the experimental and theoretical results coincide. The elastic analogs are useful to better understand several features of one-dimensional models dealt with in the quantum theory of solids. In particular, the wave amplitudes can be measured, which is only partially possible for microscopic systems.

%%%%%%%%%%%%%%%%%%%%%%%%%%%%%%%%%%%%%%%%%%%%
%% Sample figure:
%%
%% The option [height=...] scales the picture to the given height,
%% without it it would be printed at its nominal size
%%%%%%%%%%%%%%%%%%%%%%%%%%%%%%%%%%%%%%%%%%%%
%\begin{figure}
%  \includegraphics[height=.3\textheight]{golfer}
%  \caption{Picture to fixed height}
%\end{figure}
%%%%%%%%%%%%%%%%%%%%%%%%%%%%%%%%%%%%%%%%%%%%%%%%
%% BACKMATTER
%%%%%%%%%%%%%%%%%%%%%%%%%%%%%%%%%%%%%%%%%%%%%%%%

\begin{theacknowledgments}
 This work was supported by DGAPA-UNAM under projects IN111308, IN119509 and by CONACyT under projects 79613 and 82474.
\end{theacknowledgments}



\bibliographystyle{aipproc}   % if natbib is available

\end{document}